\documentclass[12pt]{article}
\pdfoutput=1
\usepackage{jheppub}
\usepackage{amsmath}
\usepackage{amsfonts}
\usepackage{amssymb}
\usepackage{graphicx}

\usepackage[export]{adjustbox}
\newcommand{\bmat}{\left(\begin{array}}
\newcommand{\emat}{\end{array}\right)}

\def\yzero{\smash{\hbox{$y\kern-4pt\raise1pt\hbox{${}^\circ$}$}}}

\def\a{\alpha}

\def\beq{\begin{equation}}
\def\eeq{\end{equation}}
\def\beqa{\begin{eqnarray}}
\def\eeqa{\end{eqnarray}}

\def\-{\hphantom{-}}
\def\ov{\overline}
\def\s2{\frac{1}{\sqrt2}}

\def\beq{\begin{equation}}
\def\eeq{\end{equation}}
\def\beqa{\begin{eqnarray}}
\def\eeqa{\end{eqnarray}}
\def\tr{{\rm tr \,}}

\def\IF{\relax{\rm I\kern-.18em F}}
\def\II{\relax{\rm I\kern-.18em I}}

\def\Dsl{\,\raise.15ex\hbox{/}\mkern-13.5mu D} 

\def\IS{{\bf {S}}}
\def\IR{{\bf {R}}}
\def\IZ{{\bf {Z}}}
\def\IX{{\bf {X}}}
\def\IY{{\bf {Y}}}

\def\IT{{\bf {T}}}
\def\IP{{\bf {P}}}

\catcode`\@=11   
\newdimen\@rotdimen
\newbox\@rotbox  

\def\@vspec#1{\special{ps:#1}}
\def\@rotstart#1{\@vspec{gsave currentpoint currentpoint translate
   #1 neg exch neg exch translate}}
\def\@rotfinish{\@vspec{currentpoint grestore moveto}}
\def\@rotr#1{\@rotdimen=\ht#1\advance\@rotdimen by\dp#1%
   \hbox to\@rotdimen{\hskip\ht#1\vbox to\wd#1{\@rotstart{90 rotate}%
   \box#1\vss}\hss}\@rotfinish}
\def\@rotl#1{\@rotdimen=\ht#1\advance\@rotdimen by\dp#1%
   \hbox to\@rotdimen{\vbox to\wd#1{\vskip\wd#1\@rotstart{270 rotate}%
   \box#1\vss}\hss}\@rotfinish}%
\def\@rotu#1{\@rotdimen=\ht#1\advance\@rotdimen by\dp#1%
   \hbox to\wd#1{\hskip\wd#1\vbox to\@rotdimen{\vskip\@rotdimen
   \@rotstart{-1 dup scale}\box#1\vss}\hss}\@rotfinish}%
\def\@rotf#1{\hbox to\wd#1{\hskip\wd#1\@rotstart{-1 1 scale}%
   \box#1\hss}\@rotfinish}%
\def\rotate{\@ifnextchar[{\@rotate}{\@rotate[l]}}
\def\@rotate[#1]#2{\setbox\@rotbox=\hbox{#2}\@nameuse{@rot#1}\@rotbox}

\catcode`\@=12

\begin{document}

\makeatletter
\@addtoreset{equation}{section}
\makeatother
\renewcommand{\theequation}{\thesection.\arabic{equation}}
\pagestyle{empty}
\rightline{IFT-UAM/CSIC-21-31}
\vspace{1.2cm}
\begin{center}
\Large{\bf
Dynamical Tadpoles,  Stringy Cobordism,\\
and the SM from Spontaneous Compactification}
\\[12mm] 

\large{Ginevra Buratti, Matilda Delgado,  Angel M. Uranga\\[4mm]}
\footnotesize{Instituto de F\'{\i}sica Te\'orica IFT-UAM/CSIC,\\[-0.3em] 
C/ Nicol\'as Cabrera 13-15, 
Campus de Cantoblanco, 28049 Madrid, Spain}\\ 
\footnotesize{\href{mailto:ginevra.buratti@uam.es}{ginevra.buratti@uam.es}, \href{mailto:matilda.delgado@uam.es}{matilda.delgado@uam.es},   \href{mailto:angel.uranga@csic.es}{angel.uranga@csic.es}}

\vspace*{5mm}

\small{\bf Abstract} \\
\end{center}
\begin{center}
\begin{minipage}[h]{\textwidth}
We consider string theory vacua with tadpoles for dynamical fields and uncover universal features of the resulting spacetime-dependent solutions. We argue that the solutions can extend only a finite distance $\Delta$ away in the spacetime dimensions over which the fields vary, scaling as $\Delta^n\sim {\cal T}$ with the strength of the tadpole ${\cal T}$. We show that naive singularities arising at this distance scale are physically replaced by ends of spacetime, related to the cobordism defects of the swampland cobordism conjecture and involving stringy ingredients like orientifold planes and branes, or exotic variants thereof. We illustrate these phenomena in large classes of examples, including AdS$_5\times T^{1,1}$ with 3-form fluxes, 10d massive IIA, M-theory on K3, the 10d non-supersymmetric $USp(32)$ strings, and type IIB compactifications with 3-form fluxes and/or magnetized D-branes. We also describe a 6d string model whose tadpole triggers spontaneous compactification to a semirealistic 3-family MSSM-like particle physics model.
\end{minipage}
\end{center}
\newpage
\setcounter{page}{1}
\pagestyle{plain}
\renewcommand{\thefootnote}{\arabic{footnote}}
\setcounter{footnote}{0}

\tableofcontents

\vspace*{1cm}

\newpage

\section{Introduction and Conclusions}

Supersymmetry breaking string vacua (including 10d non-supersymmetric strings) are generically  affected by tadpole sources for dynamical fields, unstabilizing the vacuum \cite{Fischler:1986ci,Fischler:1986tb}. We refer to them as {\em dynamical tadpoles} to distinguish them from {\em topological tadpoles}, such as RR tadpoles, which lead to topological consistency conditions on the configuration (note however that dynamical tadpoles were recently argued in \cite{Mininno:2020sdb} to relate to violation of swampland constraints of quantum gravity theories). Simple realizations of dynamical tadpoles arose in early models of supersymmetry breaking using antibranes  in type II (orientifold) compactifications \cite{Sugimoto:1999tx,Antoniadis:1999xk,Aldazabal:1999jr,Uranga:1999ib}, or in 10d non-supersymmetric string theories \cite{Polchinski:1998rr}. 

Dynamical tadpoles indicate  the fact that equations of motion are not obeyed in the proposed configuration, which should be modified to a spacetime-dependent solution (more precisely, solution in which some fields do not preserve the maximal symmetry in the corresponding spacetime dimension, but we stick to the former nomenclature), e.g. rolling down the slope of the potential. This approach has been pursued in the literature (see e.g. \cite{Dudas:2000ff,Blumenhagen:2000dc,Dudas:2002dg,Dudas:2004nd,Mourad:2016xbk}), although the resulting configurations often contain metric singularities or strong coupling regimes, which make their physical interpretation difficult. 

In this work we present large classes of spacetime\footnote{Actually, we restrict to configurations of fields varying over spatial dimensions (rather than time); yet we abuse language and often refer to them as spacetime-dependent.} dependent field configurations sourced by dynamical tadpoles, which admit a simple and tractable smoothing out of such singularities. Remarkably, these examples reveal a set of notable physical principles and universal scaling behaviours. We argue that the presence of a dynamical tadpole implies the appearance of ends of spacetime (or walls of nothing) at a finite spacetime distance, which is (inversely) related to the strength of the tadpole. These ends of spacetime moreover correspond to cobordism defects (or end of the world branes) of the theory implied by the swampland cobordism conjecture \cite{McNamara:2019rup,Montero:2020icj}. In most setups the cobordism defects end up closing off the space into a compact geometry (possibly decorated with branes, fluxes or other ingredients), thus triggering spontaneous compactification.

We can sum up the main features described above, and illustrated by our examples, in two lessons:

\smallskip

$\bullet$ {\em {\bf{\em Finite Distance:}} In the presence of a dynamical tadpole controlled by an order parameter ${\cal{T}}$, the spacetime-dependent solution of the equations of motion cannot be extended to spacetime distances beyond a critical value $\Delta$ scaling inversely proportional to ${\cal T}$, with a scaling relation}
\beqa
\Delta^{-n}\sim {\cal T}\ .
\label{bound-lesson1}
\eeqa
In our examples, $n=1$ or $n=2$ for setups with an underlying AdS-like or Minkowski vacuum, respectively.

\smallskip

$\bullet$ {\bf {\em Dynamical Cobordism:}} {\em The physical mechanism cutting off spacetime dimensions at scales bounded by the $\Delta$ above, is a cobordism defect of the initial theory (including the dynamical tadpole source).}

\smallskip

To be precise, when there are multiple spacetime directions to be closed off, the actual defect is the cobordism defect corresponding to circle or toroidal compactifications of the initial theory, with suitable monodromies on non-trivial cycles. This is analogous to the mechanism by which F-theory on half a $\IP_1$ provides the cobordism defect for type IIB on $\IS^1$ with $SL(2,\IZ)$ monodromy \cite{McNamara:2019rup} (see also \cite{Dierigl:2020lai}).

As explained, we present large classes of models illustrating these ideas, including (susy and non-susy) 10d string theories and type II compactifications with D-branes, orientifold planes, fluxes, etc. For simplicity, we present models based on toroidal examples (and orbifolds and orientifolds thereof), although many of the key ideas easily extend to more general setups. This strongly suggests that they can apply to general string theory vacua. Very remarkably, the tractability of the models allows to devise spontaneous compactification whose endpoint corresponds to some of the (supersymmetric extensions of the) SM-like D-brane constructions in the literature. As will be clear, our examples can often be regarded as novel reinterpretations of models in the literature. 

Although our examples are often related to supersymmetric models, supersymmetry is not a crucial ingredient in our discussion. Dynamical tadpoles correspond to sitting on the slope of potentials, which, even in theories admitting supersymmetric vacua, correspond to non-supersymmetric points in field space. On the other hand, supersymmetry of the final spacetime-dependent configuration is a useful trick to guarantee that dynamical tadpoles have been solved, but it is possible to build solutions with no supersymmetry but equally solving tadpoles.

\medskip

Our results shed new light on several features observed in specific examples of classical solutions to dynamical tadpoles, and provide a deeper understanding of the appearance of singularities, and the stringy mechanism smoothing them out and capping off dimensions to yield dynamical compactification. In particular, we emphasize that our discussion unifies several known phenomena and sheds new light on the strong coupling singularities of type I' in \cite{Polchinski:1995df} and in heterotic M-theory \cite{Witten:1996mz} (and its lower bound on the 4d Newton's constant). There are several directions which we leave for future work, for instance:

$\bullet$ As is clear from our explicit examples, many constructions of this kind can be obtained via a reinterpretation of known compactifications. This strongly suggests that our lessons have a general validity in string theory. It would be interesting to explore the discussion of tadpoles, cobordism and spontaneous compactifications in general setups beyond tori.

$\bullet$ A general consequence of (\ref{bound-lesson1}) is a non-decoupling of scales between the geometric scales controlling the order parameter of the dynamical tadpole and the geometric size of the spontaneously compactified dimensions. This is reminiscent of the swampland AdS distance conjecture \cite{Lust:2019zwm}. It would be interesting to explore the generation of hierarchies between the two scales, possibly based on discrete $\IZ_k$ gauge symmetries as in \cite{Buratti:2020kda}.

$\bullet$ Our picture can be regarded as belonging to the rich field of swampland constraints on quantum gravity \cite{Vafa:2005ui} (see \cite{Brennan:2017rbf,Palti:2019pca,vanBeest:2021lhn} for reviews). It would be interesting to study the interplay with other swampland constraints. In particular, the relation between the strength of the dynamical tadpole and the size of the spacetime dimensions is tantalizingly reminiscent of the first condition on $|\nabla V|/V$ of the de Sitter conjecture \cite{Obied:2018sgi,Garg:2018reu,Ooguri:2018wrx}, with ${\cal T}=|\nabla V|$ and if we interpret $V$ as the inverse Hubble volume and hence a measure of size or length scale in the spacetime dimensions. It would be interesting to explore cosmological setups and a possible role of horizons as alternative mechanisms to cut off spacetime. Also, the inequality admittedly works in different directions in the two setups, thus suggesting they are not equivalent, but complementary relations.

$\bullet$ It would be interesting to apply our ideas to the study of other setups in which spacetime is effectively cut off, such as the capping off of the throat in near horizon NS5-branes due to strong coupling effects, or the truncation in \cite{Montero:2015ofa} of throats  of the euclidean wormholes in pure Einstein+axion theories \cite{Giddings:1987cg}.

$\bullet$ Finally, we have not discussed time-dependent backgrounds\footnote{For classical solutions of tadpoles involving time dependence, see e.g. \cite{Dudas:2002dg}.}. These are obviously highly interesting, but their proper understanding is likely to require new ingredients, such as {\em end (or beginning) of time} defects (possibly as generalization of the spacelike S-branes \cite{Kruczenski:2002ap,Jones:2004rg}).

\medskip

Until we come back to these questions in future work, the present paper is organized as follows. In Section \ref{sec:conifold} we reinterpret the Klebanov-Strassler (KS) warped throat supported by 3-form fluxes as a template illustrating our two tadpole lessons. Section \ref{sec:conifold-tadpole} explains that the introduction of RR 3-form flux in type IIB theory on AdS$_5\times T^{1,1}$ produces a tadpole. The varying field configuration is the Klebanov-Tseytlin solution, which leads to a metric singularity at a finite distance scaling as (\ref{bound-lesson1}), as we show in section \ref{sec:conifold-singu}. In section \ref{sec:conifold-cobordism} we relate the KS smoothing of this singularity with cobordism defects. In section \ref{sec:conifold-generalized} we extend the discussion to other warped throats. In Section \ref{sec:3form-flux} we present a similar discussion in toroidal compactifications with fluxes. Section \ref{sec:3form-flux-tadpole} introduces a $\IT_5$ compactification with RR 3-form flux, whose tadpole backreacts producing singularites at finite distance as we show in section \ref{sec:3form-flux-singu}. In section \ref{sec:3form-flux-cobordism} we argue they are smoothed out by capping off dimensions and triggering spontaneous compactification. In Section \ref{sec:magnetization} we build examples in the context of magnetized D-branes. In section \ref{sec:magnetization-local} we describe the tadpole backreaction and its singularities, which are removed by spontaneous compactification in section \ref{sec:magnetization-global}. In Section \ref{sec:10d} we turn to the dilaton tadpole of several 10d strings. In section \ref{sec:massive-iia} we consider massive type IIA theory, where the running dilaton solutions produce dynamical cobordisms by introduction of O8-planes as cobordism defects of the IIA theory, eventually closely related to type I' compactifications. In section \ref{sec:hw} we discuss a similar picture for M-theory on K3 with $G_4$ flux, and a Horava-Witten wall as its cobordism defect. In section \ref{sec:sugimoto}  we consider the 10d non-supersymmetric $USp(32)$ theory, in two different approaches. In section \ref{sec:sugimoto-dm} we build on the classical solution in \cite{Dudas:2000ff} and discuss its singularities in the light of the cobordism conjecture. In section \ref{sec:sugimoto-magnetization} we describe an explicit (and remarkably, supersymmetry preserving) configuration solving its tadpole via magnetization and spontaneous compactification on $\IT^6$. In Section \ref{sec:the-sm} we discuss an interesting application, describing a 6d model with tadpoles, which upon spontaneous compactification reproduces a semi-realistic MSSM-like brane model. Finally, Appendix \ref{app:tadpole-wgc} discusses the violation of swampland constraints of type IIB on AdS$_5\times T^{1,1}$ when its tadpole is not duly backreacted, in a new example of the mechanism in \cite{Mininno:2020sdb}.

\section{The fluxed conifold: KS solution as spontaneous cobordism}
\label{sec:conifold}

In this section we consider the question of dynamical tadpoles and their consequences in a particular setup, based on the gravity dual of the field theory of D3-branes at a conifold singularity. The discussion is a reinterpretation, in terms useful for our purposes, of the construction of the Klebanov-Tseytlin (KT) solution \cite{Klebanov:2000nc} and its deformed avatar, the Klebanov-Strassler (KS) solution \cite{Klebanov:2000hb}. This reinterpretation however provides an illuminating template to discuss dynamical tadpoles in other setups in later sections.

We consider type IIB on AdS$_5\times T^{1,1}$, where $T^{1,1}$ is topologically $\IS^2\times \IS^3$ \cite{Klebanov:1998hh}. This is the near horizon geometry of D3-branes at the conifold singularity \cite{Klebanov:1998hh} (see also \cite{Morrison:1998cs,Uranga:1998vf,Dasgupta:1998su}), which has been widely exploited in the context of holographic dualities. The vacuum is characterized by the IIB string coupling $e^\phi=g_s$ and the RR 5-form flux $N$. The model has no scale separation, since the $T^{1,1}$ and AdS$_5$ have a common scale $R$, given by
\beqa
R^4\,=\, 4\pi\, g_sN\alpha'{}^2\ .
\label{ads-radius}
\eeqa
In any event, we will find useful to discuss the model, and its modifications, in terms of the (KT) 5d effective theory introduced in \cite{Klebanov:2000nc}. This is an effective theory not in the Wilsonian sense but in the sense of encoding the degrees of freedom surviving a consistent truncation. In particular, it includes the dilaton $\phi$ (we take vanishing RR axion for simplicity), the NSNS axion $\Phi=\int_{\IS^2}B_2$ and the $T^{1,1}$ breathing mode $q$ (actually, stabilized by a potential arising from the curvature and the 5-form flux), which in the Einstein frame enters the metric as
\beqa
&&ds^2_{10} = R^2 \big( \, e^{- 5q}\, ds^2_5 + \, e^{3q}ds^2_{T^{1,1}} \big)\ .  
\label{metric-ansatz-eft}
 \eeqa
This approach proved useful in \cite{Buratti:2018xjt} in the discussion of the swampland distance conjecture \cite{Ooguri:2006in} in configurations with spacetime-dependent field configurations (see \cite{Lust:2019zwm} for a related subsequent development, and \cite{Baume:2020dqd,Perlmutter:2020buo}).

\subsection{The 5d tadpole and its solution}
\label{sec:conifold-tadpole}

Let us introduce $M$ units of RR 3-form flux in the $\IS^3$, namely
\beqa
F_3\,=\, M\, \omega_3\ ,
\eeqa
where $\omega_3$ is defined in eq. (27) in \cite{Klebanov:2000hb}. We do not need its explicit expression, it suffices to say that it describes a constant field strength density over the $\IS^3$. The introduction of this flux sources a backreaction on the dilaton and the metric, namely a dynamical tadpole for $\phi$ and $q$. In addition, as noticed in \cite{Buratti:2018xjt}, it leads to an axion monodromy  potential for $\Phi$ \cite{Silverstein:2008sg,Kaloper:2008fb,Marchesano:2014mla,Hebecker:2014eua}. The situation is captured by the KT effective action (with small notation changes) for the 5d scalars $\phi$, $\Phi$ and $q$, collectively denoted by $\varphi^a$
\beqa
 S_{ 5} 
= - \frac{2}{\kappa_{5}^2}  \int d^{5} x 
\ \sqrt{-g_{5}} \,\bigg[ 
{ \frac 1 4}  R_5  
- { \frac 1  2} G_{ab}(\varphi)  \partial \varphi^a \partial \varphi^b 
- V(\varphi)\bigg],
\label{ktaction}
\eeqa
with the kinetic terms and potential given by 
\beqa
&& G_{ab}(\varphi)  \partial \varphi^a \partial \varphi^b \, =\,  15 (\partial q)^2  + { \frac{1}{4}}(\partial \phi )^2  + { \frac{1}{4}}  e^{ -\phi- 6 q}(\partial \Phi )^2 \ , \label{kinetic}\\
&& V(\varphi) \, = \,-5 e^{ - 8 q} + {\frac{1}{8}} M^2\, e^{\phi- 14 q}
 + {\frac{1}{8}} (N +  M \Phi)^2 e^{ - 20 q}\ . \label{potential}
\eeqa
Clearly $g_s M^2$ is an order parameter of the corresponding dynamical tadpole. In the following we focus on the case\footnote{This implies that the configuration is uncharged under a discrete $\IZ_M$ symmetry, measured by $N$ mod $M$, and associated to the redundancy generated by transformation $\phi\to \phi+1$, $N\to N-M$, see footnote \ref{foot:non-multiple}.\label{foot:multiple}} of $N$ being a multiple of $M$. 

Ignoring the backreaction of the dynamical tadpole (i.e. considering constant profiles for the scalars over the 5d spacetime) is clearly incompatible with the equations of motion. Furthermore, as argued in \cite{Mininno:2020sdb}, it can lead to violations of swampland constraints. In particular, since the introduction of $F_3$ breaks supersymmetry, if the resulting configuration was assumed to define a stable vacuum, it would violate the non-susy AdS conjecture \cite{Ooguri:2016pdq}; also, as we discuss in Appendix \ref{app:tadpole-wgc}, it potentially violates the Weak Gravity Conjecture \cite{ArkaniHamed:2006dz}. 

Hence, we are forced to consider spacetime-dependent scalar profiles to solve the equations of motion. Actually, this problem was tackled in \cite{Klebanov:2000hb}, with the scalars running with $r$, as we now review in the interpretation in  \cite{Buratti:2018xjt}. There is a non-trivial profile for the axion $\Phi$, given by 
\beqa
\Phi\, =\, 3g_sM\, \log (r/r_0)\ .
\eeqa
This implies the cancellation of the dilaton tadpole, which can be kept constant $e^{\phi}=g_s$, as follows from its equation of motion from (\ref{kinetic}), (\ref{potential})
\beqa
\nabla \phi\sim -e^{-6q-\phi}(\partial\Phi)^2+e^{-14q+\phi}M^2\ .
\eeqa

\subsection{Singularity at finite distance}
\label{sec:conifold-singu}

The varying $\Phi$ corresponds to the introduction of an NSNS 3-form flux in the configuration
\beqa
H_3\,=\, -g_s \,*_{6d} F_3\ ,
\eeqa
where the 6d refers to $T^{1,1}$ and the AdS$_5$ radial coordinate $r$, and the Hodge duality is with the AdS$_5\times T^{1,1}$ metric. This is precisely such that the complexified flux combination $G_3=F_3-\tau H_3$ satisfies the imaginary self duality (ISD) constraint making it compatible with 4d Poincar\'e invariance in the remaining 4d coordinates (and in fact, it also preserves supersymmetry). The backreaction on the metric thus has the structure in \cite{Dasgupta:1999ss,Giddings:2001yu}. The metric (\ref{metric-ansatz-eft}) takes the form
\beqa
ds_{10}^{\, 2}\,=\,  Z^{-\frac 12} \eta_{\mu\nu} dx^\mu dx^\nu\, +\,Z^{\frac 12}\, \, (\, dr^2 \, +\, r^2 ds_{T^{1,1}}^2 \, )\ ,
\eeqa
where $Z$ obeys a Laplace equation in AdS$_5$, sourced by the fluxes, and reads
\beqa
Z(r)\,=\,\frac{1} {4r^{4}}\,(g_s M)^2\, \log (r/r_0)\ .
\eeqa
The warp factor also enters in the RR 5-form flux, which decreases with $r$ as
\beqa
N(r)\,=\, \int_{\IS^5} F_5\,=\, g_sM^2 \log (r/r_0)\ .
\label{running-flux}
\eeqa
This matches nicely with the monodromy for the axion $\Phi$ as it runs with $r$ \cite{Buratti:2018xjt}.
These features (as well as some other upcoming ones) were nicely explained as the gravity dual of a Seiberg duality cascade in \cite{Klebanov:2000hb}. 

This 5d running solution in \cite{Klebanov:2000nc} solves the dynamical tadpole, but is not complete, as it develops a metric singularity at $r=r_0$. This is a physical singularity at finite distance in spacetime, whose parametric dependence on the parameters of the initial model  is as follows
\beqa
& \Delta(r) = &\int_{r_0}^r Z(r)^{\frac 14}\, dr\sim  \int_{r_0}^r (g_sM)^{\frac 12} \big[ \log (r/r_0)\big]^{\frac 14}\,\frac{dr}r \nonumber \\
&& \sim  (g_sM)^{\frac 12}  \big[ \log (r/r_0)\big]^{\frac 54}= (g_s N)^{\frac 14}\,\frac {N}{g_sM^2} \sim R\, \frac {N}{g_sM^2}\ . 
\eeqa
In the last equalities we used (\ref{running-flux}), (\ref{ads-radius}). Hence, starting with an AdS$_5\times T^{1,1}$ theory with $N$ units of RR 5-form flux, the introduction of $M$ units of RR 3-form flux leads to a breakdown of the corresponding spacetime-dependent solution at a distance scaling as $\Delta\sim M^{-2}$. Recalling that the dynamical tadpole is controlled by an order parameter ${\cal T}=g_sM^2$, this  precisely matches the scaling relation (\ref{bound-lesson1}) of the Finite Distance Lesson.

\subsection{Dynamical cobordism and the KS solution}
\label{sec:conifold-cobordism}

As is well known, the singularity in the KT solution is smoothed out in the KS solution \cite{Klebanov:2000hb}. This is given by a warped version of the deformed conifold metric, instead of the conical conifold singularity, with warp factor again sourced by an ISD combination of RR 3-form flux on $\IS^3$ and NSNS 3-form flux on $\IS^2$ times the radial coordinate. At large $r$ the KS solution asymptotes to the KT solution, but near $r\sim r_0$, the solutions differ and the KT singularity is replaced by the finite size $\IS^3$ of the deformed conifold. 

Hence, the Finite Distance Lesson still applies even when the singularity is removed, and the impossibility to extend the coordinate $r$ to arbitrary distances is implemented by a smooth physical end of spacetime. The purpose of this section is to highlight a novel insight on the KS solution, as a non-trivial realization of the swampland cobordism conjecture \cite{McNamara:2019rup,Montero:2020icj}. The latter establishes that any consistent quantum gravity theory must be trivial in (a suitably defined version of) cobordism. Namely in an initial theory given by an $n$-dimensional internal compactification space (possibly decorated with additional ingredients, like branes or fluxes), there must exist configurations describing an $(n+1)$-dimensional (possibly decorated) geometry whose boundary is the initial one. The latter describes an end of the world defect (which we will refer to as the `cobordism defect') for the spacetime of the initial theory. Since the arguments about the swampland cobordism conjecture are topological, there is no claim about the unprotected properties of the cobordism defect, although in concrete examples it can preserve supersymmetry; for instance, in maximal dimensions,  the Horava-Witten boundary is the cobordism defect for 11d M-theory, and similarly the O8-plane is the cobordism defect of type IIA theory\footnote{Other 10d theories are conjectured to admit cobordism branes, but they cannot be supersymmetric and their nature is expected to be fairly exotic, and remains largely unknown. We will come back to this point in section \ref{sec:sugimoto-dm}.}.

In our setup, the initial theory is AdS$_5\times T^{1,1}$ with $N$ units of RR 5-form flux and $M$ units of RR 3-form flux on $\IS^3$. From the above discussion, it is clear that the KS solution is just the cobordism defect of this theory\footnote{Recalling footnote \ref{foot:multiple}, the case of $N$ multiple of $M$ implies the vanishing of a $\IZ_M$ charge, and allows the cobordism defect to be purely geometrical; otherwise the cobordism defect ending spacetime must include explicit D3-branes, which are the defect killing the corresponding cobordism class \cite{McNamara:2019rup}.\label{foot:non-multiple}}. The remarkable feature is that the end of spacetime is triggered dynamically by the requirement of solving the equations of motion after the introduction of the RR 3-form flux, hence it is fair to dub it dynamical cobordism. Hence, this is a very explicit illustration of the Dynamical Cobordism Lesson.

This powerful statement will be realized in many subsequent examples in later sections, and will underlie the phenomenon of spontaneous compactification, when the cobordisms close off the spacetime directions bounding them into a compact variety.

\subsection{More general throats}
\label{sec:conifold-generalized}

A natural question is the extension of the above discussion to other AdS$_5\times \IX_5$ vacua with 3-form fluxes. This question is closely related to the search for general classes of gravity duals to Seiberg duality cascades and their infrared deformations, for which there is a concrete answer if $\IX_5$ is the real base of a non-compact toric CY threefold singularity $\IY_6$, which are very tractable using dimer diagrams \cite{Hanany:2005ve,Franco:2005rj} (see \cite{Kennaway:2007tq} for a review). 

From our perspective, the result in \cite{Franco:2005fd} is that the $\IX_5$ compactification with 3-form flux $F_3$ admits a KS-like end of the world (cobordism defect\footnote{We note in passing that the regions between different throats in the multi-throat configurations \cite{Franco:2005fd} can be regarded as domain walls interpolating between two different, but bordant, type IIB vacua.}) if $\IY_6$ admits a complex deformation which replaces its conical singularity by a finite-size 3-cycle corresponding to the homology dual of the class $[F_3]$. In cobordism conjecture terms, in these configurations the corresponding global symmetry is broken, and spacetime may close off without further ado (as the axion monodromy due to the 3-form fluxes allows to eat up the RR 5-form flux before reaching the end of the world). Such complex deformations are easily discussed in terms of the web diagram for the toric threefold, as the splitting of the web diagram into consistent sub-diagrams \cite{Franco:2005fd}. Simple examples include the deformation of the complex cone over dP$_2$ to a smooth geometry, or the deformation of the complex cone over dP$_3$ to a conifold, or to a smooth geometry. 

There are however singularities (or 3-form flux assignments), for which the complex deformations are simply not available. One may then wonder about how our Dynamical Cobordism lesson applies. The answer was provided in particular examples in \cite{Berenstein:2005xa,Franco:2005zu,Bertolini:2005di}: the infrared end of the throat contains an explicit system of fractional D-branes, which in the language of the cobordism conjecture kill the corresponding cobordism classes, and allow the spacetime to end. As noticed in these references, the system breaks supersymmetry, and in \cite{Franco:2005zu} it was moreover noticed (as later revisited in \cite{Intriligator:2005aw}) to be unstable and lead to a runaway behaviour for the field blowing up the singularity. Hence, this corresponds to an additional dynamical tadpole, requiring additional spacetime dependence, to be solved. Simple examples include the complex cone over dP$_1$, and the generic $Y^{p,q}$ theories. We will not enter the discussion of possible mechanisms to stabilize these models, since following \cite{Buratti:2018onj} they are likely to require asymptotic modifications of the warped throat ansatz (i.e. at all positions in the radial direction, including the initial one).

\section{Type IIB fluxes and spontaneous compactification}
\label{sec:3form-flux}

In this Section we construct an explicit 5d type IIB model with a tunable dynamical tadpole, and describe the spacetime-dependent solution solving its equations of motion, which is in fact supersymmetry preserving. The configuration displays dynamical cobordism resulting in spontaneous compactification to 4d. The resulting model is a simple toroidal compactification with ISD NSNS and RR 3-form fluxes \cite{Dasgupta:1999ss,Giddings:2001yu}, in particular it appeared in \cite{Marchesano:2004xz,Marchesano:2004yq}. With this perspective in hindsight, one can regard this section as a reinterpretation of the latter flux compactification. Our emphasis is however in showing the interplay of the dynamical tadpoles in the 5d theory and the consequences in the spacetime configuration solving them.

\subsection{The 5d tadpole and its solution}
\label{sec:3form-flux-tadpole}

Consider type IIB on $\IT^5$, which for simplicity we consider split as $\IT^2\times\IT^2\times \IS^1$. We label the coordinates of the $\IT^2$'s as $(x^1,y^1)$ and $(x^2,y^2)$, with periodicity 1, and introduce complex coordinates as $z^1=x^1+\tau_1 y^1$, $z^2=x^2+ \tau_2 y^2$. We also use a periodic coordinate $x^3\simeq x^3+1$ to parametrize the $\IS^1$. For simplicity, we do not consider moduli deviating from this rectangular structure\footnote{As usual, they can be removed in orbifold models, although we will not focus on this possibility.}, and also take the $\IT^5$ to have an overall radius $R$,
\beqa
ds^2\,=\, R^2\, [\, (dz^1)^2\,+\, (dz^2)^2\, +\, (dx^3)^2\,]\, .
\eeqa
The result so far is a standard 5d supersymmetric $\IT^5$ compactification.

We introduce a non-trivial dynamical tadpole source by turning on an RR 3-form flux (using conventions in \cite{Giddings:2001yu})
\beqa
F_3\,=\, (2\pi)^2\alpha'\, N\, dx^1\,dx^2\,dx^3\, .
\eeqa
The introduction of this flux does not lead to RR topological tadpoles, but induces dynamical tadpoles for diverse fields. In the following we focus on the dynamics of the 5d light fields $R$, $\tau_1$, $\tau_2$, the dilaton $\phi$ and the NSNS axion $\Phi$ defined by
\beqa
B_2\,=\, \Phi \,dy^1\,dy^2\,.
\eeqa

The discussion of the dynamical tadpole is similar to the $T^{1,1}$ example in Section \ref{sec:conifold}, so we sketch the result. There is a dilaton tadpole, arising from the dimensional reduction of the 10d kinetic term for the 3-form flux, 
\beqa
\nabla^2\, \phi\, =\, \frac{1}{12}\, e^\phi\, (F_3)^2\ .
\eeqa
Since $(F_3)^2$ is a constant source density, which does not integrate to zero over $\IT^5$, there is no solution for this Laplace equation if we assume the solution to be independent of the 5d spacetime coordinates. One possibility would be to allow for 5d spacetime dependence of $\phi$ (at least in one extra coordinate, as in \cite{Dudas:2000ff}). Here we consider a different possibility, which is to let the NSNS axion $\Phi$ acquire a dependence on one of the 5d coordinates, which we denote by $y$, as follows
\beqa
\Phi \,=\, -\,(2\pi)^2\alpha'\, \frac{N}{t_3}\, y\quad \Rightarrow\, \quad H_3\, = -\,(2\pi)^2\alpha'\, \frac{N}{t_3}\, dy^1\,dy^2\,dy\,.
\label{toroidal-axion-solution}
\eeqa
We have thus turned on NSNS 3-form field strength in the directions $y^1$, $y^2$ in $\IT^5$ and the 5d spacetime coordinate $y$. Here the sign has been introduced for later convenience, and $t_3$ is a positive real parameter allowing to tune the field strength density, whose meaning will become clear later on.

Including this new source, the dilaton equation of motion becomes
\beqa
\nabla^2\, \phi\, =\, \frac{1}{12}\, \left[\,e^\phi\, (F_3)^2\,-\,e^{-\phi}\, (H_3)^2\,\right]\, .
\eeqa
Hence, the spacetime-dependent profile (\ref{toroidal-axion-solution}) can cancel the right hand side and solve the dilaton tadpole when
\beqa
e^{2\phi}\, (F_3)^2\,=\,(H_3)^2\, .
\label{isd-square}
\eeqa
We can thus keep the dilaton constant $e^{\phi}=g_s$. Taking for simplicity purely imaginary $\tau_1=it_1$ and $\tau_2=it_2$, the condition (\ref{isd-square}) is simply
\beqa
g_s\, t_1\, t_2\, t_3\, =\, 1\ .
\label{flux-moduli-condition}
\eeqa

In addition to the dilaton, the 3-form fluxes backreact on the metric and other fields, which we discuss next.

\subsection{The singularities}
\label{sec:3form-flux-singu}

We now discuss the backreaction on the metric and other fields. For convenience, we use the complex coordinates $z^1$, $z^2$ and $z^3=x^3+iy$. In terms of these, we can write the combination
\beqa
G_3\,=\, F_3 -\tau H_3\,=\, \frac{(2\pi)^2}{4}\alpha'\, N\, (\, 
d{\ov {z}_1}\,dz_2\,dz_3+ dz_1d{\ov{z}_2}dz_3 + dz_1dz_2d{\ov{z}_3} + d{\ov{z}_1}d{\ov{z}_2}d{\ov{z}_3}\,) \quad .
\label{complex-flux}
\eeqa
Regarding $\IT^5\times \IR^1_y$ as a (non-compact) CY, this is a combination of $(2,1)$ and $(0,3)$ components, which is thus ISD. There is a backreaction on the metric and RR 4-form of the familiar black 3-brane kind. In particular, the metric includes a warp factor $Z$
\beqa
ds_{10}^{\, 2}\,=\,  Z^{-\frac 12} \eta_{\mu\nu} dx^\mu dx^\nu\, +\,Z^{\frac 12}\, R^2\, [ \, dz^1d{\ov z}^1\, +\,dz^2d{\ov z}^2\,+\,dz^3d{\ov z}^3
\,]\ ,
\eeqa
where $x^\mu$ runs through the four Poincar\'e invariant spacetime coordinates. The warp factor is determined by the Laplace equation
\beqa
-{\tilde \nabla}^2 Z = \frac{g_s}{12}G_3\cdot {\ov G}_3\, =\, \frac{g_s}6\, (F_3)^2\,,
\label{laplace-toroidal}
\eeqa
with the tilde indicating the Laplacian is computed with respect to the unwarped, flat metric, and in the last equation we used (\ref{isd-square}).

Note that, since $y$ parametrizes a non-compact dimension, there is no tadpole problem in solving (\ref{laplace-toroidal}) i.e. we need not add background charge. One may then be tempted to conclude that this provides a 5d spacetime-dependent configuration solving the 5d tadpole. However, the solution is valid locally in $y$, but cannot be extended to arbitrary distances in this direction. Since the local flux density in $\IT^5$ is constant, we can take $Z$ to depend only on\footnote{In fact, this is the leading behaviour at long distances, compared with the $\IT^5$ size scale $R$.} $y$, hence leading to a solution
\beqa
-\frac{d^2 Z}{dy^{\,2}}\,=\,  \frac{g_s}6\, (F_3)^2\quad \Rightarrow \quad
Z\,=\, 1 \, -\, \frac{g_s}{12} (F_3)^2\, y^2\, ,
\eeqa
where we have set an integration constant to 1. The solution hits metric singularities at
\beqa
y^{-2}\,=\, \frac{1}{12}   g_s(F_3)^2\, ,
\eeqa
showing there is a maximal extent in the direction $y$. Let us introduce the quantity ${\cal T}=\frac{1}{12}   g_s(F_3)^2$, which controls the parametric dependence of the tadpole. Then, the distance between the singularities is
\beqa
\Delta\,=\, \int_{-{\cal T}^{-1/2}}^{{\cal T}^{-1/2}} \, Z^{\frac14}\, dy\, =\, \frac 2{\sqrt{{\cal T}}}\, \int_0^1 (1-t^2)^{\frac 14}\, dt\ ,
\eeqa
with $t=\sqrt{{\cal T}}y$. We thus recover the scaling (\ref{bound-lesson1}) with $n=2$,
\beqa
\Delta^{-2}\sim {\cal T}\ .
\label{scaling}
\eeqa
Hence the appearance of the singularities as a consequence of the dynamical tadpole is as explained in the introduction.

\subsection{Cobordism and spontaneous compactification}
\label{sec:3form-flux-cobordism}

The appearance of singularities is a familiar phenomenon. In this section we argue that they must be smoothed out, somewhat analogously to the KS solution in section \ref{sec:conifold}. The fact that it is possible follows from the swampland cobordism conjecture \cite{McNamara:2019rup,Montero:2020icj}, namely there must exist an appropriate cobordism defect closing off the extra dimension into nothing. 
Since there are two singularities, the formerly non-compact dimension becomes compact, in an explicit realization of spontaneous compactification\footnote{Spontaneous compactification has been discussed in the context of dynamical tadpoles in \cite{Dudas:2000ff}.}.

In the following, we directly describe the resulting geometry, which turns out to be a familiar  $\IT^6$ (orientifold) compactification with ISD 3-form fluxes. Consider type IIB theory on $\IT^2\times \IT^2\times \IT^2$, with
\beqa
F_3\,=\, (2\pi)^2\alpha'\, N\, dx^1\,dx^2\,dx^3\ , \quad H_3\,=\, (2\pi)^2\alpha'\, N\, dy^1\,dy^2\,dy^3\ .
\eeqa
We use $z^i=x^i+it_i y^i$, hence the above defined $t_3$ is the complex structure modulus for the $\IT^2$ involving the newly compact dimension. For moduli satisfying (\ref{flux-moduli-condition}) the $\IT^6$ flux combination $G_3$ is given by (\ref{complex-flux}), which is ISD and indeed compatible with 4d Poincar\'e invariance as usual. Notice that in this case, it is possible to achieve a large size for the new compact dimension $t_3\gg 1$ by simply e.g. taking small $g_s$. This corresponds to the regime of small 5d tadpole, with the relation
\beqa
t_3^{-2}\sim g_s^{\,2}\sim \mathcal{T}^2\ ,
\eeqa
in agreement with the maximal distance relation in the previous section.

Consistency, in the form of $C_4$ RR tadpole cancellation, requires the introduction of O3-planes at fixed points of the involution ${\cal R}:z^i\to -z^i$ (together with mobile D3-branes). From the perspective of the 5d theory, the additional dimension is compactified on an interval, with two end of the world defects given by the O3-planes, which constitute the cobordism defects of the configuration (possibly decorated with explicit D3-branes if needed).

\section{Solving dynamical tadpoles via magnetization}
\label{sec:magnetization}

In this Section we consider a further setup displaying dynamical tadpoles, based on compactifications with magnetized D-branes \cite{Bachas:1995ik,Angelantonj:2000hi,Blumenhagen:2000wh,Aldazabal:2000dg,Aldazabal:2000cn}. In toroidal setups, these have been (either directly or via their T-dual intersecting brane world picture) widely used to realize semi-realistic particle physics models in string theory. In more general setups, magnetized 7-branes are a key ingredient in the F-theory realization of particle physics models \cite{Donagi:2008ca,Beasley:2008dc,Beasley:2008kw}.

\subsection{Solving dynamical tadpoles of magnetized branes}
\label{sec:magnetization-local}

We consider a simple illustrative example. Consider type IIB theory compactified on $\IT^2 \times \IT^2$ (labelled 1 and 2, respectively) and mod out by $\Omega {\cal R}_1(-1)^{F_L}$, where ${\cal R}_1:z_1\to -z_1$. This introduces  4 O7$_1$-planes spanning $(\IT^2)_2$ and localized at the fixed points on $(\IT^2)_1$. We also have 32 D7-branes (as counted in the covering space), split as 16 D7-branes (taken at generic points) and their 16 orientifold images. This model is related by T-duality on $(\IT^2)_1$ to a type I toroidal compactification, but we proceed with the D7-brane picture.

We introduce $M$ units of worldvolume magnetic flux along $(\IT^2)_2$ for the $U(1)$ of a D7-brane\footnote{If $N$ the D7-branes are coincident, it is also possible to use the overall $U(1)\subset U(N)$. We will stick to the single D7-brane for the moment, but such generalization will arise in later examples.\label{foot-u1}}
\beqa
\frac{1}{2\pi \alpha'}\,\int_{\IT^2} F_2\,=\, M\, .
\label{magnetic-flux-quant}
\eeqa
The orientifold requires we introduce $-M$ units of flux on the image D7-brane\footnote{For simplicity we consider vanishing discrete NSNS 2-form flux \cite{Blumenhagen:2000ea}, although such generalization will arise in later examples.}. This also ensures that there is no net induced $\IZ$-valued D5-brane charge in the model, and hence no associated RR tadpole, in agreement with the fact that the RR 6-form is projected out. In addition, there is a $\IZ_2$ K-theory charge \cite{Uranga:2000xp} which is cancelled as long as $M\in 2\IZ$.

The introduction of the worldvolume flux leads to breaking of supersymmetry. As is familiar in the discussion of supersymmetries preserved by different branes \cite{Berkooz:1996km}, we introduce the angle
\beqa
\theta_2\,=\,  \arctan ( 2\pi \alpha'\, F)\,=\,  \arctan \left( M\chi \right)\ ,
\label{angle-6d}
\eeqa
where $F$ is the field strength and $\chi$ is the inverse of the $\IT^2$ area, in string units.

This non-supersymmetric configuration introduces dynamical tadpoles. For small $\theta_2$, the extra tension can be described in effective field theory as an FI term controlled by $\theta$ \cite{Kachru:1999vj,Cvetic:2001nr,Cvetic:2001tj}. In fact, in \cite{Cremades:2002te} a similar parametrization was proposed for arbitrary angles. By using the DBI action, the extra tension has the structure
\beqa
V\, \sim\, \frac {1}{g_s}\, \left(\, \sqrt{1+(\tan\theta_2)^2} -1\,\right)
\,.
\eeqa
This leads to a tadpole for the dilaton and the $(\IT^2)_2$ K\"ahler modulus.

We now consider solving the tadpole by allowing for some spacetime-dependent background. Concretely, we allow for a non-trivial magnetic field $-F$ on two of the non-compact space coordinates, parametrized by the (for the moment, non-compact) coordinate $z_3$. In fact this leads to a configuration preserving supersymmetry since, defining the angle $\theta_3$ in analogy with (\ref{angle-6d}), we satisfy the $SU(2)$ rotation relation $\theta_3+\theta_2=0$ \cite{Berkooz:1996km}. In other words, the field strength flux has the structure
\beqa
F_2\,=\, F (\,dz_2d{\ov z}_2\, - \, dz_3d{\ov z}_3\,)\ ,
\eeqa
which is $(1,1)$ and primitive (i.e. $J\wedge F_2=0$), which are the supersymmetry conditions for a D-brane worldvolume flux.

Hence, it is straightforward to find spacetime-dependent solutions to the tadpole of the higher-dimensional theory, at the price of breaking part of the symmetry of the lower-dimensional spacetime. In the following we show that, as in earlier examples, this eventually also leads to spontaneous compactification.

\subsection{Backreaction and spontaneous compactification}
\label{sec:magnetization-global}

The spacetime field strength we have just introduced couples to gravity and other fields, so we need to discuss its backreaction. 

In fact, this is a particular instance of earlier discussions, by considering the F-theory lift of the D7-brane construction. This can be done very explicitly by taking the configuration near the $SO(8)^4$ weak coupling regime \cite{Sen:1996vd}. The configuration without magnetic flux $M=0$ simply lifts to F-theory on K3$\times \IT^2\times \IR^2$, where the $(\IT^2)_1$ (modulo the $\IZ_2$ orientifold action) is the $\IP_1$ base of K3, and the $\IT^2$ and $\IR^2$ explicit factors correspond to the directions $z_2$ and $z_3$, respectively. As is familiar, the 24 degenerate fibers of the K3 elliptic fibration form 4 pairs, reproducing the 4 orientifold planes, and 16 D7-branes in the orientifold quotient. Actually, the discussion below may be carried out for F-theory on K3 at generic points in moduli space, even not close to the weak coupling point.

The introduction of magnetization for one 7-brane corresponds to the introduction of a $G_4$ flux along the local harmonic $(1,1)$-form supported at an $I_1$ degeneration (or enhanced versions thereof, for coincident objects), of the form
\beqa
G_4\,=\, \omega_2\wedge F (\,dz_2d{\ov z}_2\, - \, dz_3d{\ov z}_3\,)\ .
\eeqa
This flux is self-dual, and in fact $(2,2)$ and primitive, which is the supersymmetry preserving condition for 4-form fluxes in M/F-theory \cite{Dasgupta:1999ss,Becker:2001pm}. The backreacted metric is described by a warp factor satisfying a Laplace equation sourced by the fluxes, similar to (\ref{laplace-toroidal}). Considering the regime in which the warp factor is taken independent of the internal space and depends only on the coordinates in the $\IR^2$ parametrized by $z_3$, the constant flux density leads to singularities at a maximal length scale $\Delta$
\beqa
\Delta^{-2}\,\sim\,  F^2\, .
\eeqa 
This is another instance of the universal relation (\ref{bound-lesson1}) with ${\cal T}\sim F^2$, hence $n=2$.

This is in complete analogy with earlier examples. Hence, we are led to propose that the smoothing out of these singularities is provided by the compactification of the corresponding coordinates, e.g. on a $\IT^2$, with the addition of the necessary cobordism defects, namely orientifold planes and D-branes\footnote{To be precise, the cobordism defects of an $\IS^1$ compactification of the model. This is analogous to the mechanism by which F-theory on half a $\IP_1$ provides the cobordism defect for type IIB on $\IS^1$ with $SL(2,\IZ)$ monodromy \cite{McNamara:2019rup} (see also \cite{Dierigl:2020lai}). In fact, since magnetized branes often lead to chiral theories in the bulk, this extra circle compactification allows them to become non-chiral and admit an end of the world describable at weak coupling, see the discussion below (\ref{scaling-distance-dm}) in section \ref{sec:sugimoto}. We will nevertheless abuse language and refer as cobordism defect to the structures involved in the final spontaneous compactification under discussion.}.

To provide an explicit solution, we introduce the  standard notation (see e.g. \cite{Blumenhagen:2000wh,Aldazabal:2000dg}) of $(n,m)$ for the wrapping numbers and the magnetic flux quanta on the $(\IT^2)_i$'s for the directions $i=1,2,3$. In this notation, the O7$_1$-planes and unmagnetized D7$_1$-branes are associated to $(0,1)\times (1,0)\times (1,0)$, while the magnetized D7$_1$-branes\footnote{If the magnetization is in the $U(1)\subset U(N)$ of a stack of $N$ coincident branes, see footnote \ref{foot-u1}, the corresponding wrapping goes as $(N,M)$.} correspond to $(0,1)\times (1,M)\times (1,-M)$, and $(0,1)\times (1,-M)\times (1,M)$ for the orientifold images. In other words, we require a flux quantization condition on $(\IT^2)_3$ as in (\ref{magnetic-flux-quant}), up to a sign flip.

Since now the last complex dimension is compact, there is an extra RR tadpole cancellation condition, which requires the introduction of 16 O7$_3$-planes, wrapped on $(\IT^2)_1\times (\IT^2)_2$ and localized at fixed points in $(\IT^2)_1$, namely with wrapping numbers $(1,0)\times (1,0)\times (0,1)$. This introduces an extra orbifold action generated by $(z_1,z_2,z_3)\to (z_1,-z_2,-z_3)$, so the model can be regarded as a (T-dual of a) magnetized version of the D9/D5-brane $\IT^4/\IZ_2$ orientifolds in \cite{Pradisi:1988xd,Gimon:1996rq}. Allowing for $n$ additional mobile D7$_3$-branes (as counted in the covering space, and arranged in orbifold and orientifold invariant sets), the RR tadpole cancellation conditions is
\beqa
2M^2+n = 32\ .
\eeqa
The supersymmetry condition is simply that the $\IT^2$ parameters satisfy $\chi_2=\chi_3$.

From the perspective of the original 6d configuration, the tadpole in the initial $\IT^2\times \IT^2$ configuration has triggered a spontaneous compactification. Since the additional O-planes and D-branes required to cancel the new RR tadpoles are localized in $z_3$, they can be interpreted as the addition of I-branes to cancel the cobordism charge of the original model.

\medskip

It should be possible to generalize the above kind of construction to global K3-fibered CY threefolds with O7-planes. The local fibration in a small neighbourhood of a generic point of the base provides a local 6d model essentially identical to our previous one. On the other hand, the global geometry defining how the two extra dimensions compactify would correspond to another possible spontaneous compactification, with the ingredients required for the cancellation of the new RR tadpoles. 

However, a general drawback of this class of models is that the scales of the compact spaces in the directions 2 and 3 are of the same order\footnote{In the toroidal example, if the magnetization along $(\IT^2)_2$ is on the overall $U(1)\subset U(16)$ of 16 coincident D7-branes, the magnetic field along $z_2$ is $F\sim M/16$; this weakened tadpole implies an increase of the critical size of the spontaneously compactified dimensions by a factor of 4.}. Thus, there is no separation of scales, and no reliable regime in which the dynamics becomes that of a 6d model. This is easily avoided in more involved models, as we will see in the examples in coming sections.

\section{Solving tadpoles in 10d strings}
\label{sec:10d}

In this section we consider dynamical tadpoles arising in several 10d string theories, and confirm the general picture. We illustrate this with various examples, with superymmetry (massive type IIA and M-theory on K3), and without it (non-supersymmetric 10d $USp(32)$ theory).

\subsection{Massive IIA theory}
\label{sec:massive-iia}

We consider 10d massive type IIA theory \cite{Romans:1985tz}. This can be regarded as the usual type IIA string theory in the presence of an additional RR 0-form field strength $F_0\equiv m$. The string frame effective action for the relevant fields is
\beqa
S_{10}\,=\, \frac{1}{2\kappa_{10}^{\,2}}\int d^{10}x \,\sqrt{-G}\, \{\, e^{-2\phi}\, [\, R+4(\partial \phi)^2\,]-\frac 12 (F_0)^2 -\frac 12 (F_4)^2\,\}\, +\, S_{\rm top}\, ,
\eeqa
where $S_{\rm top}$ includes the Chern-Simons terms. In the Einstein frame $G_E=e^{-\frac{\phi}{2}} G$, we have
\beqa
S_{10,E}\,=\,  \frac{1}{2\kappa^{\,2}}\int d^{10}x \,\sqrt{-G_E}\, \{\, [\, R-\frac 12(\partial \phi)^2\,]-\frac 12 e^{\frac 52\phi} m^2  -\frac 12 e^{\frac 12 \phi}(F_4)^2\,\}\, .
\eeqa
Here we have used $m$ to emphasize this quantity is constant. This theory is supersymmetric, but at a  given value of $\phi$, it has a tadpole controlled by
\beqa
{\cal T}\,\sim \, e^{\frac 52\phi} m^2\ .
\label{tadpole-massiveiia}
\eeqa
This is in particular why the massive IIA theory does not admit 10d maximally symmetric solutions. In the following we discuss two different ways of solving it, leading to Minkowski or AdS-like configurations.

\subsubsection{Solution in 9d and type I' as cobordism}
\label{sec:type-iprime}

To solve the tadpole (\ref{tadpole-massiveiia}) we can consider a well-known 1/2 BPS solution with the dilaton depending on one coordinate $x^9$. Since the flux $m$ can be regarded as generated by a set of $m$ distant D8-branes, this is closely related to the solution in \cite{Bergshoeff:1995vh}. We describe it in  conventions closer to \cite{Polchinski:1995df}, for later use. In the Einstein frame, the metric and dilaton background have the structure
\beqa
(G_{E})_{MN}\, =\, Z(x^9)^{\frac 1{12}}\,\eta_{MN}\ ,\quad
e^{\phi}\,=\, Z(x^9)^{-\frac 56}\ , \quad{\rm with}\; \;Z(x^9)\sim B-mx^9\ ,
\eeqa
where $B$ is some constant (in the picture of flux generated by distant D8-branes, it relates to the D8-brane tensions).
The solution hits a singularity at $x^9=B/m$. Starting at a general position $x^9$, the distance to the singularity is
\beqa
\Delta\,=\, \int_{x^9}^{\frac Bm} Z(x^9)^{\frac{1}{24} }\,dx^9\,\sim Z(x^9)^{\frac{25}{24}}\,m^{-1}\, \sim \,  m^{-1} e^{-\frac{5}{4} \phi}\ ,
\eeqa
where in the last equality we have traded the position for the value  the dilaton takes there.
Recalling (\ref{tadpole-massiveiia}), this reproduces the Finite Distance scaling relation (\ref{bound-lesson1}) with $n=2$,
\beqa
\Delta^{-2}\sim {\cal T}\ .
\eeqa
It is easy to propose the stringy mechanism capping off spacetime before or upon reaching this singularity, according to the Dynamical Cobordism lesson. This should be the cobordism defect of type IIA theory, which following \cite{McNamara:2019rup} is an O8-plane, possibly with D8-branes. 

In fact, this picture is implicitly already present in \cite{Polchinski:1995df}, which studies type I' theory, namely type IIA on an interval, namely IIA on $\IS^1$ modded out by $\Omega{\cal R}$ with ${\cal R}:x^9\to -x^9$, which introduces two O8$^-$-planes which constitute the interval boundaries. There are 32 D8-branes (in the covering space), distributed on the interval, which act as domain walls for the flux $F_0=m$, which is piecewise constant in the interval. The metric and the dilaton profile are controlled by a piecewise linear function $Z(x^9)$. The location of the boundaries at points of strong coupling was crucial to prevent contradiction with the appearance of certain enhanced symmetries in the dual heterotic string (the role of strong coupling at the boundaries for the enhancements was also emphasized from a different perspective in \cite{Seiberg:1996bd,Kachru:1996nd}). In our setup, we interpret the presence of (at least, one) O8-plane as the cobordism defect triggered by the presence of a dynamical tadpole in the bulk theory.

\subsubsection{A non-supersymmetric Freund-Rubin solution}
\label{sec:freund-rubin}

We now consider for illustration a different mechanism to cancel the dynamical tadpole, which in fact underlies the spontaneous compactification to (non-supersymmetric\footnote{Thus, it should be unstable according to \cite{Ooguri:2016pdq}. However, being at a maximum of a potential is sufficient to avoid dynamical tadpoles, so the  solution suffices for our present purposes.}) AdS$_4\times \IS^6$ in \cite{Romans:1985tz}). The idea is that, rather than solving for the dilaton directly, one can introduce an additional flux $F_4$ along  three space dimensions and time (or its dual $F_6$ on six space dimensions) to balance off the dilaton sourced by $F_0$. This can be used to fix $\phi$ to a constant, and following  \cite{Romans:1985tz} leads to a scaling
\beqa
F_4\sim \,m^2\, d(vol)_4\, ,
\eeqa
where $d(vol)_4$ is the volume form in the corresponding 4d. Using arguments familiar by now, the constant $F_4$ backreaction on the metric is encoded in a solution of the 4d Laplace equation with a constant source, leading to a solution quadratic in the coordinates (to avoid subtleties, we take solutions depending only on the space coordinates). This develops a singularity at a distance scaling as
\beqa
\Delta^2\sim |F_4|^{-2}\sim m^{-4}\sim {\cal T}\ ,
\eeqa
where in comparison with (\ref{tadpole-massiveiia}) we have taken constant dilaton. 

The singularities are avoided by an AdS$_4\times \IS^6$ compactification, whose curvature radius is $R\sim m^2$, in agreement with the above scaling. From our perspective, the compactification should be regarded as a dynamical cobordism (where the cobordism is actually that of the 10d theory on an $\IS^5$ (i.e. equator of $\IS^5$)).

\subsection{An aside on M-theory on K3}
\label{sec:hw}

In this section we relate the above system to certain compactifications of M-theory and  to the Horava-Witten end of the world branes as its cobordism defect. Although the results can be obtained by direct use of M-theory effective actions, we illustrate how they can be recovered by applying simple dualities to the above system.

Consider the above massive IIA theory with mass parameter $m$, and compactify on $\IT^4/\IZ_2$. This introduces O4-planes, and requires including 32 D4-branes in the configuration, either as localized sources, or dissolved as instantons on the D8-branes. Actually this can be considered as a simple model of K3 compactifications, where in the general K3 the O4-plane charge is replaced by the contribution to the RR $C_5$ tadpole arising from the CS couplings of D8-branes and O8-planes to $\tr R^2$. 

We now perform a T-duality in all the directions of the $\IT^4/\IZ_2$ (Fourier-Mukai transform in the case of general K3). We obtain a similar model of type I' on $\IT^4/\IZ_2$, but now with the tadpole being associated to the presence of $m$ units of non-trivial flux of the RR 4-form field-strength over $\IT^4$ (namely, K3). Also, the dilaton of the original picture becomes related to the overall K\"ahler modulus of K3. Finally, we lift the configuration to M-theory by growing an extra $\IS^1$ and decompactifying it. We thus end up with a 7d compactification of M-theory on K3, with $m$ units of $G_4$ flux,
\beqa
\int_{\rm K3} G_4\,=\,m\ .
\eeqa
This leads to a dynamical tadpole, cancelled by the variation of the overall K\"ahler modulus (i.e. the K3 volume) along one the 7d space dimensions, which we denote by $x^{11}$. As in previous sections, this will trigger a singularity at a finite distance in $x^{11}$, related to the tadpole by $\Delta^{-2}\sim {\cal T}$. The singularity is avoided by the physical appearance of a cobordism defect, which for M-theory is a Horava-Witten (HW) boundary \cite{Horava:1995qa}. This indeed can support the degrees of freedom to kill the $G_4$ flux, as follows. From \cite{Horava:1996ma}, the 11d $G_4$ is sourced by the boundary as
\beqa
dG_4\,=\, \delta(x^{11})\, (\, \tr F^2\,-\,\frac 12 \tr R^2\,)\ ,
\eeqa
where $\delta(x^{11})$ is a bump 1-form for the HW brane, and $F$ is the field-strength for the $E_8$ gauge fields in the boundary. Hence, the $m$ units of $G_4$ in the K3 compactification can be absorbed by a HW boundary with an $E_8$ bundle with instanton number $12+m$ (the 12 coming from half the Euler characteristic of K3 $\int_{\rm K3} \tr R^2=24$).

The above discussion is closely related to the picture in \cite{Witten:1996mz}, which discusses compactification of HW theory (namely, M-theory on an interval with two HW boundaries) on K3 and on a CY threefold. It includes a K\"ahler modulus varying over the interval according to a linear function\footnote{In the presence of explicit M5-branes, it is a piecewise linear function. It is straightforward to include them in our cobordism description if wished, with explicit branes considered as part of the cobordism defect.} and the appearance of a singularity at finite distance. In that case, the HW brane was located at the strong coupling point, based on heuristic arguments, and this led, in the CY$_3$ case, to a lower bound on the value of the 4d Newton's constant. 

Our perspective remarkably explains that the location of the HW wall is not an arbitrary choice, but follows our physical principle of Dynamical Cobordism, and the bound on the Newton's constant is a consequence of that of Finite Distance!

\subsection{Solving tadpoles in the non-supersymmetric 10d \texorpdfstring{$USp(32)$}{USP(32)} theory}
\label{sec:sugimoto}

The previous examples were based on an underlying supersymmetric vacuum, on top of which the dynamical tadpole is generated via the introduction of fluxes or other ingredients. In this section we consider the opposite situation, in which the initial theory is strongly non-supersymmetric and displays a dynamical tadpole from the start. In particular we consider the non-supersymmetric 10d $USp(32)$ theory constructed in \cite{Sugimoto:1999tx}, in two different ways: first, we use our new insights to revisit the spacetime-dependent solution proposed in \cite{Dudas:2000ff} (see also \cite{Blumenhagen:2000dc} for other proposals); then we present a far more tractable solution involving magnetization, which in fact provides a supersymmetric compactification of this non-supersymmetric 10d string theory.

\subsubsection{The Dudas-Mourad solution and cobordism}
\label{sec:sugimoto-dm}

The non-supersymmetric 10d $USp(32)$ theory in \cite{Sugimoto:1999tx} is obtained as an $\Omega$ orientifold of type IIB theory. The closed string sector is as in type I theory, except that the O9$^-$-plane is replace by an O9$^+$-plane. Cancellation of RR tadpoles requires the introduction of open strings, which must be associated to 32 ${\ov{\rm D9}}$-branes. The closed string sector is a 10d ${\cal N}=1$ supergravity multiplet; the orientifold action on the D9-branes breaks supersymmetry, resulting in an open string sector with $USp(32)$ gauge bosons and gauginos in the two-index antisymmetric representation. All anomalies cancel, a remarkable feat from the field theory viewpoint, which is just a consequence of RR tadpole cancellation from the string viewpoint.

Although the RR tadpoles cancel, the NSNS tadpoles do not, implying that there is no maximally symmetric 10d solution to the equations of motion. In particular there is a dynamical dilaton tadpole of order the string scale, as follows from the terms in the 10d (Einstein frame) action 
\beqa
S_E\,=\, \frac{1}{2\kappa^2}\int d^{10}x \sqrt{-G}\,[\, R-\frac{1}{2}(\partial\phi)^2\,] \,-\, T_9^E\int d^{10}x \sqrt{-G}\, 64\, e^{\frac{3\phi}2}\, ,
\label{action-sugimoto}
\eeqa
where $T_9^E$ is the (anti)D9-brane tension. The tadpole scales as ${\cal T}\sim T_9^E g_s^{3/2}$, with the dilaton dependence arising from the fact that the supersymmetry breaking arises from the Moebius strip worldsheet topology, with $\chi=3/2$. 

Ref. \cite{Dudas:2000ff} proposed solutions of this dynamical tadpole with 9d Poincar\'e invariance, and the dilaton varying over one spacetime dimension (see also \cite{Antonelli:2019nar,Basile:2020mpt} for more recent, related work). In the following we revisit the solution with dependence on one spatial coordinate $y$, from the vantage point of our Lessons.

The 10d solution is, in the Einstein frame,
\beqa
\phi &=& \frac 34 \alpha_E y^2\,+\,\frac 23\log|\sqrt{\alpha_E}y|\,+\,\phi_0 \ ,\nonumber\\
ds_E^{\,2}&=& |\sqrt{\alpha_E}y|^{\frac 19}\, e^{-\frac{\alpha_E y^2}8} \eta_{\mu\nu} dx^\mu dx^\nu +  |\sqrt{\alpha_E}y|^{-1} e^{-\frac{3\phi_0}2} e^{-\frac{9\alpha_E y^2}{8}}\, dy^2\, ,
\label{solution-dm}
\eeqa
where $\alpha_E= 64 k^2T_9$. There are two singularities, at $y=0$ and $y\to\infty$, which despite appearances are separated by a finite distance 
\beqa
\Delta\,\sim\, \int_0^\infty  |\sqrt{\alpha_E}y|^{-\frac 12} e^{-\frac{3\phi_0}4} e^{-\frac{9\alpha_E y^2}{16}}\, dy\,\sim \,e^{-\frac{3\phi_0}4} \alpha_E^{-\frac 12}\ .
\eeqa
The fact that the solution has finite extent in the spatial dimension on which the fields vary is in agreement with the Finite Distance Lesson, and in fact satisfying its quantitative bound (\ref{bound-lesson1})
\beqa
\Delta^{-2}\sim {\cal T}\ .
\label{scaling-distance-dm}
\eeqa

We can now consider how the Dynamical Cobordism Lesson applies in the present context. Following it, we expect the finite extent in the spatial dimensions to be physically implemented via the cobordism defect corresponding to the 10d $USp(32)$ theory. In general the cobordism defect of bulk chiral 10d theories are expected to be non-supersymmetric, and in fact rather exotic, as their worldvolume dynamics must gap a (non-anomalous) set of chiral degrees of freedom. In fact, on general grounds they can be expected to involve strong coupling\footnote{We are indebted to Miguel Montero for this argument, and for general discussions on this section.}. An end of the world defect imposes boundary conditions on bulk supergravity fields, which at weak coupling should be at most linear in the fields, to be compatible with the superposition principle. A typical example are boundary conditions that pair up bulk fermions of opposite chiralities. However, the anomaly cancellation in the 10d $USp(32)$ theory involves fields of different spins, which cannot be gapped by this simple mechanism, and should require strong coupling dynamics (a similar phenomenon in a different context occurs in \cite{Razamat:2020kyf}).

This strong coupling fits nicely with the singularity at $y\to \infty$, but the singularity at $y=0$ lies at weak coupling. The simplest way out of this is to propose that the singularity at $y=0$ is actually smoothed out by perturbative string theory (namely, $\alpha'$ corrections, just like orbifold singularities are not singular in string theory), and does not turn into an end of the world defect. Hence the solution (\ref{solution-dm}) extends to $y<0$, and, since the background is even in $y$, develops a singularity at $y\to -\infty$. This is still at finite distance $\Delta$ scaling as (\ref{scaling-distance-dm}), and lies at strong coupling, thus allowing for the possibility that the singularity is turned into the cobordism defect of the 10d $USp(32)$ theory.

It would be interesting to explore this improved understanding of this solution to the dilaton tadpole. Leaving this for future work, we turn to a more tractable solution in the next section.

\subsubsection{Solving the tadpole via magnetization}
\label{sec:sugimoto-magnetization}

We now discuss a more tractable alternative to solve the dynamical tadpole via magnetization, following section \ref{sec:magnetization}.

Stabilizing the tadpole via magnetization is, ultimately, equivalent to finding a compactification (on a product of $\IT^2$'s) which is free of tadpoles, for instance by demanding it to be supersymmetric. Hence we need to construct a supersymmetric compactification of the non-supersymmetric 10d $USp(32)$ theory \cite{Sugimoto:1999tx}. 

As explained above, the 10d model is constructed with an O9$^+$-plane and 32 ${\ov{\rm D9}}$-branes. Hence, we need to introduce worldvolume magnetic fields in different 2-planes, in such a way that the corresponding angles add up to 0 mod $2\pi$. It is easy to convince oneself that this requires  magnetization in at least three complex planes, ultimately triggering a $\IT^2\times \IT^2\times \IT^2$ compactification. In order to preserve supersymmetry, we need the magnetization to induce D5-brane charges, rather than ${\ov{\rm D5}}$-brane charge, hence we need the presence of three independent kinds of negatively charged O5$^-_i$-planes, where $i=1,2,3$ denotes the $\IT^2$ wrapped by the corresponding O5-plane. We are thus considering an orientifold of $\IT^6/(\IZ_2\times\IZ_2)$ with an O9$^+$-plane, and 8 O5$^-_i$-planes\footnote{For such combinations of orientifold plane signs, see the analysis in \cite{Klein:2000qw}, in particular its table 6. We will not need its detailed construction for our purposes.}.

The wrapping numbers for the O-planes, and for one simple solution of all constraints for the D9-branes (and their explicitly included orientifold image D9-branes), are

%
\begin{center}
\begin{tabular}{|c|c||c|c|c|}
\hline
Object & $N_\a$   &  $(n_\a^{1},m_\a^{1})$ &  $(n_\a^{2},m_\a^{2})$ & $(n_\a^{3},m_\a^{3})$\\
\hline\hline 
O9$^+$ & $32$ &  $(1,0)$ & $(1,0)$  & $(1,0)$\\
\hline 
O5$^-_1$ & $-32$ &  $(1,0)$ & $(0,1)$  & $(0,-1)$\\
\hline 
O5$^-_2$ & $-32$ &  $(0,1)$ & $(1,0)$  & $(0,-1)$\\
\hline 
O5$^-_3$ & $-32$ &  $(0,1)$ & $(0,-1)$  & $(1,0)$\\
\hline 
D9 & $16$ &  $(-1,1)$ & $(-1,1)$  & $(-1,1)$\\
\hline 
D9$'$ & $16$ &  $(-1,-1)$ & $(-1,-1)$  & $(-1,-1)$\\
\hline \end{tabular}
\end{center}

It obeys the RR tadpole conditions for the $\IZ$-valued D9- and D5-brane charges, and the discrete $\IZ_2$ RR tadpole conditions for D3- and D7$_i$-brane charges \cite{Uranga:2000xp}.

The supersymmetry condition determined by the O-plane wrappings is
\beqa
\sum_i\arctan (-\chi_i)\equiv \theta_1+\theta_2+\theta_3=0\; {\rm mod}\; 2\pi \, .
\eeqa
The model is in fact T-dual (in all $\IT^6$ directions) to that in section 5 of \cite{Marchesano:2004xz}.

It is easy to see that the above condition forces at least one of the $\IT^2$ to have ${\cal O}(1)$ area in $\alpha'$ units. From our perspective, this a mere reflection of the fact that the 10d dynamical tadpole to be canceled is of order the string scale, hence it agrees with the scaling $\Delta^{-2}\sim {\cal T}$. Happily, the use of an $\alpha'$ exact configuration, which is moreover supersymmetric, makes our solution reliable. This is an improvement over other approaches e.g. as in section \ref{sec:sugimoto-dm}.

Although we have discussed the compactification on (an orientifold of) $\IT^6$ directly, we would like to point out that it is easy to describe it as a sequence of $\IT^2$ spontaneous compactifications, each eating up a fraction of the initial 10d tadpole until it is ultimately cancelled upon reaching $\IT^6$. However, this picture does not really correspond to a physical situation, given the absence of decoupling of scales. This is true even in setups which seemingly allow for one $\IT^2$ of parametrically large area. Indeed, consider for instance the regime $\chi_3\sim 2\lambda$ and $\chi_1,\chi_2\sim \lambda^{-1}$, for $0<\lambda\ll 1$, which corresponds to $\theta_1,\theta_2\sim \frac \pi{2}+\lambda$, $\theta_3\sim \pi- 2\lambda$. This corresponds to a compactification on substringy size $(\IT^2)_1\times (\IT^2)_2$  and a parametrically large $(\IT^2)_3$. However, the fact that the $(\IT^2)_1$, $ (\IT^2)_2$ can be T-dualized into large area geometries shows that there is not true decoupling of scales: in the original picture, the small sizes imply that there are towers of light winding modes, whose scale is comparable with the KK modes of $(\IT^2)_3$. Hence, the lack of decoupling is still present, as expected from our general considerations in the introduction.

\section{The SM from spontaneous compactification}
\label{sec:the-sm}

In this section we explore an interesting application of the above mechanism, and provide an explicit example of a 6d theory with brane-antibrane pairs, and a dynamical tadpole triggering spontaneous compactification to a 4d (MS)SM-like particle physics model. Interestingly, the complete chiral matter and electroweak sector, including the Higgs multiplets, are generated as degrees of freedom on cobordism branes. Only the gluons are present in some form in the original 6d models.

Consider the type IIB orientifold of $\IT^4/\IZ_2$ with orientifold action $\Omega$ constructed in \cite{Pradisi:1988xd,Gimon:1996rq}, possibly with magnetization. To describe it,  we introduce the notation in \cite{Blumenhagen:2000ea,Ibanez:2001nd}  of wrapping numbers $(n_\alpha^i,m_\alpha^i)$, where $n_\alpha^i$ and  $m_\alpha^i$ provide the wrapping number and magnetic flux quantum of the D-brane $\alpha$ on the $i^{\rm th}$ $\IT^2$, respectively. We consider the following stacks of D-branes (and their orientifold images, not displayed explicitly)

\smallskip

%
\begin{center}
\begin{tabular}{|c||c|c|}
\hline
 $N_\a$   &  $(n_\a^{1},m_\a^{1})$ &  $(n_\a^{2},m_\a^{2})$ \\
\hline\hline $N_{a+d} = 6+2$ &  $(1,3)$ & $(1,-3)$  \\
\hline
\hline $N_{h_1}= 4$ &  $(1,-3)$ & $(1,-4)$ \\
\hline $N_{h_2}= 4$ & $(1,-4)$ & $(1,-3)$ \\
\hline $40 $ &   $(0,1)$  & $(0,-1)$  \\
\hline \end{tabular}
\end{center}

%
\smallskip

The O9- and O5-planes correspond to the wrapping numbers $(1,0) \times (1,0)$ and $(0,1)\times (0,-1)$ respectively.
The stacks $a$ and $d$ are taken different and separated by Wilson lines, but they can be discussed jointly for the time being. They correspond to 8 D9-branes with worldvolume magnetic fluxes 72 units of D5-brane charge.  The stacks $h_1$ and $h_2$ correspond to 8 additional D9-branes, with 96 with  units of induced  ${\ov{\rm D5}}$-branes charge. The addition of 40 explicit D5-branes leads to RR tadpole cancellation (once orientifold images are included).  In terms of the wrapping numbers, we have
\beqa
&\sum_\alpha N_\alpha n_\alpha^2n_\alpha^3=16
\quad , \quad
&\sum_\alpha N_\alpha m_\alpha^2m_\alpha^3=-16\ .
\eeqa

The model is far from supersymmetric due to the presence of D5-${\ov{\rm D5}}$ pairs, and in fact has a decay channel to supersymmetric model by their annihilation. On the other hand, even at the top of the tachyon potential, the theory is not at a critical point of its potential due to dynamical tadpole for the closed string moduli, namely the area moduli of the $\IT^2$'s. In other words, the excess tension depends on these, as they enter the angles determining the deviation from the supersymmetry condition
\beqa
\arctan  \Big( \frac {m_\alpha^1}{n_\alpha^1}\chi_1\Big)\, +\, \arctan  \Big( \frac {m_\alpha^2}{n_\alpha^2}\chi_2\Big)\,=\,0\, .
\eeqa
For instance, we can make the stacks $a$, $d$ supersymmetric, by choosing
\beqa
\chi_1=\chi_2\ ,
\eeqa
but the D-branes $h_1$ and $h_2$ break supersymmetry. Hence, there is a dynamical tadpole associated to the excess tension of these latter objects.

The dynamical tadpole can be solved by introducing magnetization along two of the 6d spacetime dimensions. The backreaction of this extra flux forces these two dimensions to be compactified on a $\IT^2$, with the addition of cobordism I-branes \cite{Montero:2020icj}, which in general includes orientifold planes and D-branes, as in the examples above. We take these extra branes to be  arranged in two new stacks $b$ and $c$. Overall, we end up with an orientifold of $\IT^6/(\IZ_2\times\IZ_2)$, with D-brane stacks and topological numbers given by
\smallskip

%
\begin{center}
\begin{tabular}{|c||c|c|c|}
\hline
 $N_\a$   &  $(n_\a^{1},m_\a^{1})$ &  $(n_\a^{2},m_\a^{2})$ & $(n_\a^{3},m_\a^{3})$\\
\hline\hline $N_{a+d} = 6+2$ &  $(1,3)$ & $(1,-3)$  & $(1,0)$\\
\hline $N_b=2$ &   $ (0,1)$  & $(1,0)$  & $(0,1)$ \\
\hline $N_c=2$ &  $(-1,0)$  & $(0,-1)$  & $(0,1)$\\
\hline
\hline $N_{h_1}= 2$ &  $(1,-3)$ & $(1,-4)$ & $(2,-1)$\\
\hline $N_{h_2}= 2$ & $(1,-4)$ & $(1,-3)$ & $(2,-1)$\\
\hline $40 $ &   $(0,1)$  & $(0,-1)$  & $(0,1)$\\
\hline \end{tabular}
\end{center}

The model satisfies the RR tadpole conditions
\beqa
& \sum_\alpha\,N_\alpha n_\alpha^1n_\alpha^2n_\alpha^3\,=\, 16\ ,\quad & \sum_\alpha\,N_\alpha n_\alpha^1m_\alpha^2m_\alpha^3\,=\, 16\ , \nonumber \\
& \sum_\alpha\,N_\alpha m_\alpha^1n_\alpha^2m_\alpha^3\,=\, 16\ ,\quad & \sum_\alpha\,N_\alpha m_\alpha^1m_\alpha^2n_\alpha^3\,=\, -16\ . 
\eeqa
This corresponds to O9-planes along $(1,0)\times (1,0)\times (1,0)$, and O5-planes along $(0,1)\times (0,-1)\times (1,0)$, as already present in the 6d theory, and cobordism O5-planes along $(0,1)\times (1,0)\times(0,1)$ and $(1,0)\times (0,1)\times (0,1)$. 

The model still contains only 3 stacks of D-branes with non-trivial angles, so that they are just enough to fix the 2 parameters $\chi_i$ of the $\IT^2$'s. The O-planes fix the supersymmetry condition signs to
\beqa
\arctan  \Big( \frac {m_\alpha^1}{n_\alpha^1}\chi_1\Big)\, +\, \arctan  \Big( \frac {m_\alpha^2}{n_\alpha^2}\chi_2\Big)\,-\, \arctan  \Big( \frac {m_\alpha^3}{n_\alpha^3}\chi_3\Big)\,=\,0\, .
\eeqa
Using the branes above, we get
\beqa
\chi_1=\chi_2\quad ,\quad \chi_3\,=\, \frac{14\chi_1}{1-12\chi_1^{\,2}}\,.
\eeqa
The regime of large $(\IT^2)_3$ corresponds to small $\chi_3$, which is also attained for small $\chi_1$. Note that in this context the last condition $\chi_1\sim \chi_3$ encodes the relation between the 6d tadpole and the inverse area of the spontaneously compactified $\IT^2$.

The model is, up to exchange of directions in the $\IT^6$ and overall sign flips, precisely one of the examples of 4d MSSM-like constructions in \cite{Marchesano:2004xz,Marchesano:2004yq}. The gauge group is $U(3)_a\times USp(2)_b\times U(1)_c \times U(1)_d$, where we break the naive $USp(2)_c$ by Wilson lines or shifting off the O-plane for the corresponding D5-branes. Taking into account the massive $U(1)$'s due to $BF$ couplings, this reproduces the SM gauge group. In addition, open strings between the different brane stacks reproduce a 3-family (MS)SM chiral matter content, and the MSSM Higgs doublet pair. Hence, we have described the spontaneous compactification of a 6d model to a  semi-realistic MSSM-like 4d theory.

A fun fact worth emphasizing is that most of the SM spectrum is absent in the original 6d model, and arises only after the spontaneous compactification. In particular, all the MSSM matter and Higgs chiral multiplets, as well as the electroweak gauge sector, arise from open string sectors involving the $b$ and $c$ branes, which arises as cobordism branes. It is remarkable that cobordism entails that spontaneous compactification implies not just the removal of spacetime dimensions, but also the dynamical appearance of novel degrees of freedom. It is tantalizing to speculate on the potential implications of these realizations in cosmological or other dynamical setups.

%
\section*{Acknowledgments}
We are pleased to thank Inaki Garc\'ia-Etxebarria, Luis Ib\'anez, Fernando Marchesano, Miguel Montero and Irene Valenzuela for useful discussions. This work is supported by the Spanish Research Agency (Agencia Espa\~nola de Investigaci\'on) through the grants IFT Centro de Excelencia Severo Ochoa SEV-2016-0597, the grant GC2018-095976-B-C21 from MCIU/AEI/FEDER, UE.

\newpage

\appendix
\section{Dynamical tadpoles and swampland constraints}
\label{app:tadpole-wgc}
In this appendix we use the model in section \ref{sec:conifold} to illustrate the result in \cite{Mininno:2020sdb} that, in theories with a dynamical tadpole which is not duly backreacted on the field configuration, the mistreatment can show up as violations of swampland constraints. 

We consider type IIB theory on AdS$_5\times T^{1,1}$ and introduce $M$ units of RR 3-form flux. In the coordinates in \cite{Klebanov:1998hh,Klebanov:2000hb}, it reads
\beqa
    F_3&=&\frac{1}{2} M [\,\sin{\theta _1} (\cos{\theta_2} d\theta_1d\phi _1 d\phi_2+d\theta_1d\phi_1d\psi)
    +\sin{\theta _2}(\cos{\theta _1}  d\theta_2 d\phi_1 d\phi_2- d\theta _2 d\phi_2 d\psi)\,]\nonumber\ .
\eeqa
It has constant coefficients in terms of {\em f\"unf}-bein 1-forms $g^i$ in \cite{Klebanov:2000hb} $F_3=\frac 12 M g^5(g^1g^2+g^3g^3)$, hence its kinetic term $|F_3|^3$ is constant over the $T^{1,1}$ geometry. This acts as a constant background source for e.g. the Laplace equation for the dilaton, which has no solution over the compact $T^{1,1}$ geometry. This inconsistency of the equations of motion, assuming no backreaction on the underlying geometry, signals the dynamical tadpole in the configuration. In the following we will argue that it moreover can lead to violation of the Weak Gravity Conjecture \cite{ArkaniHamed:2006dz}.

For concreteness we focus on the simplest set of states, corresponding to 5d BPS particle states in the original theory ($M=0$), with the BPS bound corresponding to the WGC bound, for the gauge interaction associated to the KK $U(1)$ dual to the $U(1)_R$ symmetry of the dual CFT. For small R-charge $n\ll N$, these particle states are dual to chiral primary single-trace mesonic operators of the $SU(N)^2$ theory, e.g. $\tr (A_1B_1\ldots A_1B_1)$; in the AdS side, they correspond to KK gravitons with momentum $n$ on the $\IS^1$. For very large R-charge, the KK gravitons polarize due to Myers' effect \cite{Myers:1999ps} into giant gravitons \cite{McGreevy:2000cw}, and their dual operators are determinant or sub-determinant operators \cite{Balasubramanian:2001nh}. Note that on $T^{1,1}$ we have D3-branes wrapped on homologically trivial 3-cycles (but sustained as BPS states by their motion on $\IS^1$), hence they are different from (di)baryonic operators, which correspond to D3-branes wrapped on the non-trivial $\IS^3$ \cite{Gubser:1998fp}.

Our strategy is to consider these states in the presence of $F_3$, but still keeping the geometry as AdS$_5\times T^{1,1}$ (i.e. with no backreaction of the dynamical tadpole), and show that the interaction of $F_3$ makes these states non-BPS, hence violating the WGC bound. This analysis will be quite feasible in the giant graviton regime $1\ll n\sim N$, by using the wrapped D3-brane worldvolume action. Admittedly, proving a full violation of the WGC would require showing the violation of the BPS condition for all values of $n$; we nevertheless consider the large $n$ result as a compelling indication that the WGC is indeed violated in this configuration, thus making its inconsistency manifest.

Supersymmetric 3-cycles for D3-branes are easily obtained from holomorphic 4-cycles in the underlying CY threefold \cite{Mikhailov:2000ya,Beasley:2002xv} (see also \cite{Forcella:2008au}). Describing the conifold as $z^1z^2-z^3z^4=0$, any holomorphic function of these coordinates $f(x,y,z,w)=0$ defines a holomorphic 4-cycle corresponding to a giant graviton D3-branes, i.e. wrapped on a trivial\footnote{Di-baryonic D3-branes are on the other hand associated to non-Cartier divisors in the conifold, i.e. 4-cycles which can be defined in terms of the $a_i$, $b_i$ homogeneous coordinates of the linear sigma model, but cannot be expressed as a single equation $f(z^i)=0$.} 3-cycle in $T^{1,1}$. We focus on a simple class of D3-branes studied in detail in \cite{Hamilton_2010}. They are defined by the 4-cycle $z^1=\sqrt{1-\alpha^2}$, with $\alpha\in[0,1]$ being a real constant, encoding the size of the 3-cycle (with $\alpha=0,1$ corresponding to the pointlike KK graviton and the maximal giant graviton, respectively). We will follow the analysis in \cite{Hamilton_2010} with the inclusion of the effect of $F_3$ on the D3-brane probe.

It is convenient to change to new coordinates \(\{\chi_1,\chi_2,\chi_3,\alpha,\nu\}\)
\begin{equation}
    \begin{cases}
    \chi_1= \frac{1}{3}(\psi-\phi_1-\phi_2)\\
    \chi_2=\frac{1}{3}(\psi+3\phi_1-\phi_2)\\
    \chi_3=\frac{1}{3}(\psi-\phi_1+3\phi_2)
    \end{cases} \;\; \begin{cases}
    \sqrt{1-\alpha^2}=\sin{\frac{\theta_1}{2}}\sin{\frac{\theta_2}{2}}\\
    \nu=\frac{2u}{\alpha^2+u^2}\;\; \text{with}\;\; u=\cos{\frac{\theta_1}{2}}\cos{\frac{\theta_2}{2}}
    \end{cases}
\end{equation}
These are adapted to the D3-brane embedding, which simply reads
\[ \sigma^0=t,\;\; \sigma^1=\nu \;(\text{doubly-covered}),\;\;\sigma^2=\chi_2,\;\; \sigma^3=\chi_3\ .\]
The double covering is very manifest for the maximal giant graviton, $\alpha=1$, $z^1=0$. It corresponds to the defining equation $z^3z^4=0$, which splits in two components, corresponding to two (oppositely oriented) copies of the non-trivial\footnote{In terms of the linear sigma model coordinates we have $z^1=a_1b_1$, $z^2=a_2b_2$, $z^3=a_1b_2$, $z^4=a_2b_1$, and the two components correspond to $a_1=0$ and $b_1=0$, which are non-Cartier divisors.} $\IS^3$. The double covering remains even for non-maximal giants, even though they correspond to irreducible 4-cycles.

The RR 3-form field strength in these coordinates is
\begin{equation}
\begin{gathered}
        F_3=M ([a_{12}\;d\chi_1\wedge d\chi_2+a_{13}\; d\chi_1\wedge d\chi_3+a_{23}\; d\chi_2\wedge d\chi_3]\wedge d{\alpha}\\+[v_{12}\; d\chi_1\wedge d\chi_2+v_{13}\; d\chi_1\wedge d\chi_3+v_{23}\;d\chi_2\wedge d\chi_3]\wedge d{\nu} )\ ,
\end{gathered}
\end{equation}
with
\[
\begin{cases}
a_{12}=\frac{9}{4} \alpha   (1\pm\frac{\sqrt{1-\nu ^2}}{1-c}  )\\
a_{13}=\frac{9}{4} \alpha    (-1\pm\frac{\sqrt{1-\nu ^2}}{1-c}  )\\
a_{23}=\frac{9}{4} \alpha    (\frac{-c}{1-c} )
\end{cases}
\quad
\begin{cases}
v_{12}=\mp\frac{9}{4}\frac{   c  (\nu ^2-c )}{ \nu ^3 \sqrt{1-\nu ^2} (1-c) }\\
v_{13}=\mp\frac{9}{4}\frac{   c  (\nu ^2-c )}{ \nu ^3 \sqrt{1-\nu ^2} (1-c)}\\
v_{23}=-\frac{9}{4}\frac{   c^2}{ \nu ^3 (1-c)}
\end{cases}
\] 
where we have introduced \(c=1-\sqrt{1-\alpha^2 \nu^2}\). We can fix a gauge and find the RR 2-form
\begin{equation}
    C_2= M( c_{12} d\chi_1\wedge d\chi_2 +c_{13} d\chi_1\wedge d\chi_3+c_{23} d\chi_2\wedge d\chi_3)\ ,
\end{equation}
with
\[\begin{cases}
c_{12}=-\frac{9}{8} (-\alpha ^2\mp\frac{2 \sqrt{1-\nu ^2}  c }{\nu ^2} )\\
c_{13}=-\frac{9}{8}  (\alpha ^2\mp\frac{2 \sqrt{1-\nu ^2}  c }{\nu ^2} )\\
c_{23}=\frac{9}{8} \frac{   (\alpha ^2 \nu ^2-2c )}{ \nu ^2}
\end{cases}\]
Its pullback on the D3-brane worldvolume is
\begin{equation}
 P[C_2]= M\dot\chi_1(  c_{12} dt\wedge d\chi_2 + c_{13}dt\wedge d\chi_3)+ M c_{23} d\chi_2\wedge  d\chi_3\ .
\end{equation}
We can now compute the effect of this background on the D3-brane by using its worldvolume action. This is easy in the S-dual frame, in which the RR 2-form couples to the D3-brane just like the NSNS 2-form in the original DBI+CS D3-brane action\footnote{Related to this, one can check that the above background is neither pure gauge on the D3, nor cannot be removed by a change in the worldvolume gauge field strength flux.}.
After integrating over $\chi_2$, $\chi_3$, this reads
\begin{equation}\begin{gathered}
    S=S_{BDI}+S_{CS} = \frac{64 \pi^2}{9} \int dt L\ , \\\text{with}\quad L=\int_0 ^1 d\nu\;2\, (- T_3 \sqrt{-\det{(P[G]_{\mu\nu}+P[C_2]_{\mu\nu})}}+ \mu_3 R^4 c_4 \dot \chi_1)\ ,
\end{gathered}\end{equation}
where the factor of 2 of the double-covering of \(\nu\) has been added, and the last term arises from the CS coupling to the RR 4-form as in \cite{Hamilton_2010}.

We are interested in focusing on the angular momentum of the state \(P_{\chi_1}=\frac{\partial L}{\partial \dot \chi_1}\) conjugate to the angular coordinate \(\chi_1\). This reads
\begin{equation}
    P_{\chi_1}= \frac{3}{2} \int_0 ^1d\nu  \Big(\, \frac{\sqrt{3 \pi } A g_\nu \,N T_3(\dot\chi_1)}{\sqrt{ N\,g_\nu(B-A (\dot\chi_1)^2)}}+9 \pi  c_4 \mu_3 \, N\, \Big)\ ,
\end{equation}
with
\begin{equation}
    \begin{gathered}
            A=\frac{81}{64 \nu ^4} \{\,-4 \,M^2 [\,(2 (1-\alpha ^2 \nu ^2)c-\alpha ^2 \nu ^2) c\,]
            -3 \pi  {N}(\alpha ^2-1) \nu ^2  c^2\,\}\,\equiv\, A_{M^2}{M^2}+ A_N {N}\ ,\\
            B=\frac{81}{64 \nu ^4} \big\{4 {M}^2\big[\,\big(4-\alpha ^2 \nu ^2\big)c^2-2 \alpha ^2 \nu ^2 c\,\big]\\
            +\pi {N}\big[\, 2 \alpha ^4 \nu ^4+\big(\,\alpha ^2 \nu ^2-2 c\big)(-3 \alpha ^2 \nu ^2-3 \nu ^2+8)\,\big]\,\big\}\\
             \,\equiv\, B_{M^2}{M^2}+ B_N {N}\ .\\
    \end{gathered}
\end{equation}
In the last equalities we have highlighted the parametric dependence on $N$ and $M$.

Despite the fact that we have not managed to find a closed form for the result, since $M\ll N$ we can find an expansion for the integrand in the form
\beqa
n= p_0(\alpha,\nu, \dot \chi_1)\,N+ p_2(\alpha,\nu, \dot \chi_1)\, M^2+ {\cal O}(M^4)\ ,
\eeqa
where the coefficient functions are computable, but we will not need their explicit expressions.

The coefficient $p_0$ is the survivor for the $M=0$ case, and leads to an integer momentum. On the other hand, the subleading correction $p_2$ produces a momentum which is not integer. This already signals a problem, since (as the geometry is considered undeformed even after introducing $F_3$) the gauge coupling of the KK $U(1)$ is as in the $M=0$ case, hence charges under it should be integer in the same units. Hence one can directly claim that the assumption of ignoring the dynamical tadpole backreaction lead to violation of charge quantization, in contradiction with common lore for consistency with quantum gravity \cite{Banks:2010zn}.

The above discussion however seems to contradict the fact that any quantum excitation on a periodic $\IS^1$ direction must have quantized momentum to have a well-defined wavefunction. In fact, an alternative interpretation of the above mismatch is that the D3-brane probe computation assumes a well-defined worldvolume embedding, in particular well-defined (hence classical) trajectories for the 5d particle. It is only for BPS states in supersymmetric vacua that such a computation is guaranteed to end up producing quantized momenta. The fact that our holomorphic embedding ansatz fails to do so is just a reflection that the actual integer-quantized states are {\em not} described by holomorphic equations. Since the latter condition is the one ensuring the match between the particle mass and charge, it is clear that non-holomorphic embeddings will produce larger masses for the same charge, hence violating the BPS / WGC bound.

\bibliographystyle{JHEP}
\bibliography{mybib}

\end{document}